\documentclass[preprint,epsfig,aps,showpacs]{revtex4}

\usepackage{epsfig}

\newcommand{\BV}{\left(\begin{array}{c}}
\newcommand{\EV}{\end{array}\right)}
\newcommand{\BM}{\left(\begin{array}{cc}}

\newcommand{\beq}{\begin{equation}}               
\newcommand{\eeq}{\end{equation}}                 
\newcommand{\bqry}{\begin{eqnarray}}              
\newcommand{\eqry}{\end{eqnarray}}                
\newcommand{\bqryn}{\begin{eqnarray*}}            
\newcommand{\eqryn}{\end{eqnarray*}}              

\begin{document}

\setcounter{page}{0}

\title{Majorana-Like
$\protect\protect\bbox{(j,0)\oplus(0,j)}$ Representation Spaces:\\ Construction
and Physical Interpretation
\footnotemark[1]\footnotetext
[1]{ This work was done under the auspices of the U. S. Department of
Energy. }}

\author{D. V. Ahluwalia} 
\affiliation{MP-9, MS H-846, Los Alamos National Laboratory,
Los Alamos, New Mexico 87545, USA}
\author{T. Goldman}
\affiliation{T-5, MS B-283, Los Alamos National Laboratory,
Los Alamos, New Mexico 87545, USA}
\author{M. B. Johnson}
\affiliation{MP-9, MS H-846, Los Alamos National Laboratory,
Los Alamos, New Mexico 87545, USA}
 
\date{19 July 1993}

\begin{flushright}
\vspace{-1.5in}
{LA-UR-93-2645}\\
\vspace{-0.15in}
{hep-th/9307118}\\
\vspace*{0.2in}
\end{flushright}

\begin{abstract}
We present a formalism that  extends the Majorana-construction to
arbitrary spin $(j,0)\oplus(0,j)$ representation spaces. For the
example case of spin-$1$, a wave equation satisfied by the
Majorana-like $(1,0)\oplus(0,1)$ spinors is constructed and its
physical content explored. The $(j,0)\oplus(0,j)$ Majorana-construct is
found to possess an unusual classical and quantum field theoretic
structure. Relevance of our formalism to parity violation, hadronic
phenomenologies, and grand unified field theories is briefly pointed
out.
\end{abstract}

\maketitle

\noindent{\bf 1.
Introduction}

A recent careful analysis \cite{BWW} of the $(j,0)\oplus(0,j)$ representation
space has resulted in a rather surprising conclusion that bosons and antibosons
in this representation space have {\it opposite} intrinsic parity (and as such
provides a hitherto unknown realization of a class of quantum field theories
which Bargmann, Wightman, and Wigner  had classified many years ago
\cite{EWBWW}). The physical origin of this result lies in the fact, for the
example case of spin-$1$ studied in Ref. \cite{BWW}, that in the
$(1,0)\oplus(0,1)$ representation space $C$ and $P$ {\it anticommute}. This
gives a prime motive to further investigate the $(j,0)\oplus(0,j)$
representation space. In this paper, we  study and  extend the Majorana
construction \cite{Majorana} for the $({1\over 2},0)\oplus(0,{1\over 2})$
representation space to the $(j,0)\oplus(0,j)$ representation space of
arbitrary spin. In addition to the above indicated motivation,  there is
significant theoretical \cite{Zralek,BK} and experimental \cite{neutrino}
interest in the subject of Majorana field and further investigation of the
Majorana-construct and its relationship with space-time symmetries should
provide useful insights for quantum field theories of truly neutral particles.
It should be noted that a few years ago, within the context of
Rarita-Schwinger/Bargmann-Wigner field \cite{RS}, Radescu \cite{fer} extended
the concept of the Majorana field to fermions of arbitrary spin. Recently,
Boudjema {\it et al.} \cite{bosa,bosb} generalized Radescu's work to bosons. As
is well known, and as the authors of Ref. \cite{bosb} indeed note, the massless
limit of the  RS/BW formalism has inherent difficulties for spins $j\ge {3\over
2}$.   In contrast, due to a theorem of Weinberg (see Sec. III of Ref.
\cite{Weinberg}),  the $(j,0)\oplus(0,j)$ representation spaces have
well-defined
massless limits. This has been recently confirmed explicitly in Refs.
\cite{DVAa,DVAb,DVAc,DVAd,DVAe,DVAf,MS} in various contexts.

The rest of the paper is composed as follows. In the next section we begin with
introducing necessary conventions and definitions by reviewing the
the Dirac-like
$(j,0)\oplus(0,j)$ spinors. This is followed, in Section 3, by generalizing the
concept of spin-$1\over 2$ Majorana-construction to the $(j,0)\oplus(0,j)$
representation space. Section 4 is then devoted  to the  explicit construction
of the Majorana-like $(1,0)\oplus(0,1)$ spinors. Section 5 presents the wave
equation satisfied by the Majorana-like $(1,0)\oplus(0,1)$ spinors. Section 6
constructs the associated field operator, and in Section 7 we present some
concluding remarks.

The formalism that we develop is valid for massive as well as massless
particles. To facilitate the study of the massless limit, we work in the
front-form \cite{PAMD} Weinberg-Soper  formalism \cite{SW,DS}  recently
developed in Ref. \cite{MS}.

\noindent{\bf 2.
Dirac-Like $\protect\bbox{({\bf j},0)\oplus(0,{\bf j})}$ Spinors}

The front-form \cite{PAMD}  Dirac-like
$(j,0)\oplus(0,j)$ covariant spinors \cite{MS}  in the Weinberg-Soper formalism
(in   the chiral
representation) are defined as:
\begin{equation}
\psi\{p^\mu\}\,=\,\left[
\begin{array}{c}
\phi_R(p^\mu)\\
\phi_L(p^\mu)
\end{array}
\right]\quad .\label{spinor}
\end{equation}
The argument $p^\mu$ of chiral-representation spinors will be enclosed in curly
brackets $\{\,\,\}$. The Lorentz transformation  of the front-form $(j,0)$
spinors is given \cite{MS} by
\begin{equation}
\phi_R(p^\mu)\,=\,\Lambda_R(p^\mu)\,
\phi_R({\overcirc p}^\mu)\,=\,
\exp\left( \protect\bbox{ \beta\cdot {\bf J}}\right)\phi_R({\overcirc p}^\mu)\,
\quad,
\label{r}
\end{equation}
and  the front-form $(0,j)$ spinors transform as
\begin{equation}
\phi_L(p^\mu)=
\Lambda_L(p^\mu)\,
\phi_L({\overcirc p}^\mu)=\exp\left(-\protect\bbox{\beta^\ast
\cdot {\bf J}}\right)\,
\phi_L({\overcirc p}^\mu)
\quad. \label{l}
\end{equation}
The ${\overcirc p}^\mu$ represents the front-form four momentum for a
particle at rest: ${\overcirc p}^\mu \equiv (p^+=m,
\,p^1=0,\,p^2=0,p^-=m)\,$.  The ${\bf J}$ are the standard
$(2j+1)\times(2j+1)$ spin matrices, and ${\protect\bbox \beta}$ is the
boost parameter introduced in Ref. \cite{MS}
\begin{equation}
\protect\bbox{ \beta}
\,=\, \eta\,\left(\alpha\,v^r\,,\,\,-i\,\alpha\,  v^r\,,\,\,1\right)\quad,
\end{equation}
where
$
\alpha\,=\,\left[1\,-\,\exp(-\eta)\right]^{-1}\,,
$
 $v^r\,\,=\,\,v_x\,+\,i\,v_y$ (and
$v^\ell\,\,=\,\,v_x\,-\,i\,v_y$). In terms of the front-form variable
$p^+ \equiv E+p_z$, one can show that
\begin{equation}
\cosh(\eta/2)=\Omega
\left(p^+ + m\right)\,,\,
\sinh(\eta/2)=\Omega
\left(p^+ - m\right)\,,
\end{equation}
with
$
\Omega\,=\,\left[1/ (2 m)\right]\sqrt{m/ {p^+}}\,.
$
The norm $\overline{\psi}\{p^\mu\}\,\psi\{p^\mu\}$, with
\begin{equation}
\overline{\psi}\{p^\mu\}\,\equiv\,\psi^\dagger\{p^\mu\}\,\Gamma^0\,
,\quad\Gamma^0
\,\equiv\,
\left[
\begin{array}{cc}
0&\openone\\
\openone&0
\end{array}
\right]\quad,
\end{equation}
is so chosen that {\it  in the massless limit} :
{a.} The Dirac-like $(j,0)\oplus(0,j)$
rest spinors identically vanish (there can be no massless particles at rest);
and
{b.} Only the Dirac-like $(j,0)\oplus(0,j)$ spinors associated with
$h=\pm j$ front-form helicity \cite{MS} degrees of freedom survive.

\medskip
\noindent
These requirements uniquely determine (up to a constant factor, which we choose
to be $1/{\sqrt{2}}\,$) the $(2j+1)$-element-column form of $\phi_R({\overcirc
p}^\mu)$ and $\phi_L({\overcirc p}^\mu)$ to be
\begin{equation}
\phi^R_{j}({\overcirc p}^\mu)\,=\,{m^j\over {\sqrt{2}}}
\left[
\begin{array}{c}
1\\
0\\
\vdots\\
0
\end{array}\right]
\,,\quad
\phi^R_{j-1}({\overcirc p}^\mu)\,=\,{m^j\over {\sqrt{2}}}
\left[
\begin{array}{c}
0\\
1\\
\vdots\\
0
\end{array}\right]
\,,\cdots\,\quad
\phi^R_{-j}({\overcirc p}^\mu)\,=\,{m^j\over {\sqrt{2}}}
\left[
\begin{array}{c}
0\\
0\\
\vdots\\
1
\end{array}\right]
\,\quad,\label{rests}
\end{equation}
[with similar expressions for $\phi_L({\overcirc p}^\mu)\,$] in a
representation in which $J_z$ is diagonal. The subscripts $h=
j,\,j-1,\,\cdots\,,-j$ on $\phi^R_h({\overcirc p}^\mu)$ in Eq. (\ref{rests})
refer to the front-form helicity \cite{MS} degree of freedom. The reader should
refer to Sec. 2.5 of Ref. \cite{MSM} for an alternate discussion of the
non-trivial nature (even though it appears as a ``normalization factor'')  of
the factor $m^j$  in Eq. (\ref{rests}).

\noindent{\bf 3.
Majorana-Like $\protect\bbox{({\bf j},0)\oplus(0,{\bf j})}$ Spinors}

Following Ramond's work \cite{Ramond} on spin-$1\over 2$, we define the
front-form $(j,0)\oplus(0,j)$ {\it $\theta$-conjugate spinor }
\begin{equation}
\psi^\theta\{p^\mu\}\equiv \left[
\begin{array}{c}
\left(\xi\,\Theta_{[\,j]}\right)\,\phi^\ast_L(p^\mu) \\
\left(\xi\,\Theta_{[\,j]}\right)^\ast\,\phi^\ast_R(p^\mu)
\end{array}\right]\quad,\label{c}
\end{equation}
where $\xi$ is a c-number, and $\Theta_{[\,j]}$ is the Wigner's
time-reversal operator
(see Refs.: p. 61 of \cite{Ray}, Eqs. 6.7 and 6.8 of the first reference
 in \cite{SW},
 and Ch. 26 of \cite{EW})
\begin{equation}
\Theta_{[\,j]}
\,\,{{\bf J}}\,\,\Theta_{[\,j]}^{-1} \,=\,-\,\protect\bbox{ {\bf
J}^{\,\ast}}\quad,
\end{equation}
and $^\ast$ denotes the operation of algebraic complex conjugation.
The parameter $\xi$ is fixed by imposing the constraint:
\begin{equation}
\left[\psi^\theta\{p^\mu\}\right]^\theta \,=\,\psi\{p^\mu\}\quad.
\label{cc}
\end{equation}
The time-reversal operator $\Theta_{[\,j]}$ is defined as:
$
\Theta_{[\,j]}\,=\,(-1)^{j+\sigma}\,\delta_{\sigma^\prime,\,-\sigma}\,.
$
It has the properties:
$
 \Theta_{[\,j]}^\ast\,\Theta_{[\,j]}\,=\,(-1)^{2j}\,,\,\,
\Theta_{[\,j]}^\ast\,=\,\Theta_{[\,j]}\,.
$
In the definition of $\Theta_{[\,j]}$, $\sigma$ and $\sigma'$ represent
eigenvalues of  ${\bf J}$. For $j={1\over 2}$ and $j=1$, the $\Theta_{[\,j]}$
have the explicit forms:
\begin{equation}
\Theta_{[1/2]}\,=\,\left[
\begin{array}{ccc}
0 & {}&-1 \\
1 & {} &0
\end{array}
\right]
\,,\,\,
\Theta_{[1]}\,=\,\left[
\begin{array}{ccccc}
0 &{}& 0 &{}& 1 \\
0 &{}& -1 &{}& 0 \\
1 &{}& 0 &{}& 0
\end{array}
\right]\quad.
\end{equation}
The  properties of  the Wigner's $\Theta_{[\,j]}$ operator  allow the parameter
$\xi$ involved in the definition of $\theta$-conjugation to be fixed as
$\pm\,i$ for fermions and $\pm \,1$ for bosons. However, without loss of
generality, we can ignore the {\it minus} sign [which contributes an {\it
overall} phase factor to the $\theta$-conjugated spinors
$\psi^\theta\{p^\mu\}$] and fix $\xi$ as
\begin{eqnarray}
\xi\,=\,\left\{\begin{array}{l}
i\,\,,\quad {\mbox{for fermions}} \\
1\,\,,\quad {\mbox{for bosons}}\qquad.
\end{array}\right. \label{alp}
\end{eqnarray}

The existence of  the
Majorana spinors for the $({1\over 2},0)\oplus(0,{1\over 2})$
representation space is
usually (see, e.g., p.16 of Ref. \cite{Ramond}) associated with the ``magic of
Pauli matrices,'' $\protect\bbox \sigma$. The reader may have already noticed
that
$i\,\Theta_{[1/2]}$  is identically equal to $\sigma_y$; and it is precisely
this  matrix that enters into the $CP$-conjugation of the
$({1\over 2},0)\oplus(0,{1\over 2})$ spinors.

The reason that the Majorana-like $(j,0)\oplus(0,j)$   representation
spaces, as opposed to the
$(j,0)\oplus(0,j)$ spaces spanned by
{\it Dirac-like spinors} defined by Eq. (\ref{spinor}),
can be constructed for arbitrary spins hinges upon two observations:
\medskip
{1.} Independent of spin, the front-form boosts for the
$(j,0)$ and $(0,j)$ spinors have the property that
$\left[\Lambda_R(p^\mu)
\right]^{-1}\,=\,\left[
 \Lambda_L(p^\mu)\right]^\dagger\,,\quad
\left[\Lambda_L(p^\mu)\right]^{-1}\,=\,
\left[\Lambda_R(p^\mu)\right]^ \dagger
\,;
$
and
{2.} Existence of the Wigner's time-reversal matrix
$\Theta_{[\,j]}$
 for any spin.

\medskip
\noindent
These two observations when coupled with the transformation
properties of the right- and left-handed  spinors, Eqs. (\ref{r},\ref{l}),
imply
that if $\phi_R(p^\mu)$ transforms as $(j,0)$, then
$(\zeta\,\Theta_{[\,j]})^\ast \,\phi_R^\ast(p^\mu)$ transforms as $(0,j)$
spinor. Similarly, if $\phi_L(p^\mu)$ transforms as $(0,j)$, then
$(\zeta\,\Theta_{[\,j]})^\ast\, \phi_L^\ast(p^\mu)$ transforms as $(j,0)$
spinor. Here, $\zeta=\exp(i\,\vartheta)$ is an arbitrary phase factor.  As such
we introduce $(j,0)\oplus(0,j)$ {\it Majorana-like spinors }
\begin{eqnarray}
(j,0) \,&\mapsto&\quad\rho\{p^\mu\}\,=\,
\left[
\begin{array}{c}
\phi_R(p^\mu) \\
\left(\zeta_\rho \,\Theta_{[\,j]}\right)^\ast\,\,\phi_R^\ast(p^\mu)
\end{array}\right]\quad,\nonumber\\
(0,j) \,&\mapsto&\quad \lambda\{p^\mu\}\,=\,
\left[
\begin{array}{c}
\left(\zeta_\lambda \,\Theta_{[\,j]}\right)\,\,\phi_L^\ast(p^\mu)\\
\phi_L(p^\mu)
\end{array}\right]
\,\,.\label{fb}
\end{eqnarray}
For formal reasons, the operator multiplying $\phi_L^\ast(p^\mu)$, in the
definition of $\lambda\{p^\mu\}$, is written as $\zeta_\lambda\,\Theta$ rather
than $\left(\zeta'_\lambda\,\Theta\right)^\ast$: What we have done, in fact, is
exploited the property $\Theta^\ast=\Theta$ and chosen $\zeta_\lambda =
\zeta_\lambda^{'\,\ast}$. Since $\zeta_\lambda$ is yet to be determined, this
introduces no loss of generality. The advantage of all this is that
$\rho\{p^\mu\}$ and $\lambda\{p^\mu\}$ can now be seen as nothing but Weyl
spinors (in the $2(2j+1)$-element form)
\begin{equation}
\psi_R\{p^\mu\}\,=\,\left[
\begin{array}{c}
\phi_R(p^\mu)\\
0
\end{array}
\right]
\,,\,
\psi_L\{p^\mu\}\,=\,
\left[
\begin{array}{c}
0\\
\phi_L(p^\mu)
\end{array}
\right]
\quad,
\end{equation}
{\it added} to their  respective $\theta$-conjugates. The condition (\ref{cc})
is  satisfied not only by Majorana-like self-$\theta$-conjugate
spinors but also by antiself-$\theta$-conjugate spinors. Allowing for this
freedom, we now fix $\zeta_\rho$ and $\zeta_\lambda$ by demanding
(the defining property of the Majorana-like spinors):
\begin{equation}
\rho^\theta\{p^\mu\}\,=\,\pm\,\rho\{p^\mu\}\quad
{\mbox {and}}\quad
\lambda^\theta\{p^\mu\}\,=\,\pm\,\lambda\{p^\mu\}\quad,\label{sc}
\end{equation}
and find:
\begin{equation}
\zeta_\rho\,=\,\pm\,\xi \quad
{\mbox {and}}\quad
\zeta_\lambda\,=\,\pm\,\xi\quad.\label{zeta}
\end{equation}
The choice $\zeta_\rho =\zeta_\lambda=\,+\,\xi$ yields
self-$\theta$-conjugate spinors $\rho^{S_\theta}\{p^\mu\}$
and $\lambda^{S_\theta}\{p^\mu\}$;
while $\zeta_\rho =\zeta_\lambda=\,-\,\xi$ corresponds to
antiself-$\theta$-conjugate spinors
$\rho^{A_\theta}\{p^\mu\}$
and $\lambda^{A_\theta}\{p^\mu\}$.

It may be noted that the Dirac-like spinors, Eq. (\ref{spinor}), and
the Majorana-like spinors, Eqs. (\ref{fb}), are the {\it only} spinors that
can be introduced in any $P$-covariant theory
\footnotemark[1]
\footnotetext[1]{A theory that is covariant under the operation of parity is
{\it not} necessarily a parity non-violating theory. See Sec. 7 for a brief
discussion of this point.}  in the $(j,0)\oplus(0,j)$ representation space. The
former describe particles with a conserved charge (which may be zero), while
the
latter are inherently for the description of  neutral particles.

Before we proceed further, we make a few observations on the definition of
$\theta$-conjugation. For the $({1\over 2},0)\oplus(0,{1\over 2})$ case, the
definition (\ref{c}) of $\theta$-conjugation can be verified to coincide with
$CP$-conjugation. The reader may wish to note that what Raymond (see Ref.
\cite{Ramond}, p. 20) calls a ``charge conjugate spinor,'' in the context of
spin-$1\over 2$,  is actually a $CP$-conjugate spinor. This, we suspect,
remains
true for fermions of higher spins also. Surprisingly, for the
$(1,0)\oplus(0,1)$ spinors (and presumably for bosons of higher spins also)
$\theta$-conjugation  equals $\Gamma^5\,C$ within a phase factor
\footnotemark[2]
\footnotetext[2]{ The charge conjugation
operator C, along with $P$ and $T$, for the $(1,0)\oplus(0,1)$ spinors and
fields was recently obtained in Ref. \cite{BWW}.}
of $-(-1)^{|h|}\,$.
The mathematical origin of
this fact may be traced to the constraint (\ref{cc}) and the property
$\Theta^\ast_{[j]}\,\Theta_{[j]}=(-1)^{2j}$ of the Wigner's time-reversal
operator $\Theta\,$. One may ask if one changes the  constraint (\ref{cc}) to
read $\left[\psi^\theta\{p^\mu\}\right]^\theta\,=\, -\psi^\theta\{p^\mu\}$ for
bosons, whether one can obtain an alternate definition of $\theta$-conjugation
(so that $\theta$-conjugation equals $CP$ for bosons also) to construct
self/antiself-$\theta$-conjugate objects. A simple exercise reveals that no
such construction yields self/antiself-$\theta$-conjugate objects. The reader
may wish to note parenthetically that when the
result (\ref{alp}) is coupled with
the definition of $\theta$-conjugation, Eq. (\ref{c}), we discover that the
operation of $\theta$-conjugation  treats the right-handed and  left-handed
spinors in a fundamentally asymmetric fashion for {\it fermions}. This is
readily inferred by studying the relative phases with which
$\left(\xi\,\Theta_{[\,j]}\right)\,\phi^\ast_L(p^\mu) $ and
$\left(\xi\,\Theta_{[\,j]}\right)^\ast\,\phi^\ast_R(p^\mu)$ enter in Eq.
(\ref{c}).

\noindent{\bf 4.
Explicit Construction of Majorana-Like $\protect\bbox{({\bf
1},0)\oplus(0, {\bf 1})}$ Spinors}

We now cast these formal considerations into more concrete form by studying the
$(1,0)\oplus(0,1)$ Majorana-like representation space   as an example. As in
Ref. \cite{MS}, we introduce
\footnotemark[3]
\footnotetext[3]{The generalized canonical representation
is introduced here for no other reason except to be able to compare the
results of the present work with our earlier work of Ref. \cite{MS}.}
the generalized canonical representation in the
front form:
\begin{equation}
\psi[\,p^\mu]\,=\,
{1\over{\sqrt{2}}}\,\left[
\begin{array}{ccc}
\openone &{\,\,\,}& \openone\\
\openone &{\,\,\,}& -\openone
\end{array}
\right]\,
\psi\{p^\mu\}\quad.\label{s}
\end{equation}
The argument $p^\mu$ of  canonical-representation spinors will be enclosed in
square brackets $[\,\,]$. Here $\openone$ is the $(2j+1)\times(2j+1)$ identity
matrix. The boost $M(p^\mu)$, which connects the rest-spinors
$\psi[\,{\overcirc p}^\mu]$ with the spinors associated with front-form
four momentum $p^\mu$, $\psi[\,p^\mu]$, is determined from Eqs. (\ref{r}),
(\ref{l}), and (\ref{s}):
\begin{equation}
\psi[\,p^\mu]\,=\,
M(p^\mu)\,
\psi[\,{\overcirc p}^\mu]\,,\quad
M(p^\mu)\,=\,
\left[
\begin{array}{ccc}
{\cal A}&{\,\,}& {\cal B}\\
{\cal B}&{\,\,}& {\cal A}
\end{array}
\right]
\quad,
\end{equation}
with
$
{\cal A}= \Lambda_R(p^\mu)
+\Lambda_L(p^\mu)\,,\,\,
{\cal B}=
\Lambda_R(p^\mu)
-\Lambda_L(p^\mu)\,.
$

Using the identities (needed to evaluate $M(p^\mu)$ explicitly) given in Ref.
\cite{MS} we first obtain the spin-$1$ $\rho^{S_\theta}[\,p^\mu]$. These are
tabulated in Table I. The $\rho^{S_\theta}[\,p^\mu]$ spinors satisfy the
following orthonormality relations: $ \overline\rho^{\,S_\theta}
_{h}[\,p^\mu]\,\rho^{S_\theta}
_{h'}[\,p^\mu]\,=\, m^2\Theta_{h h'} \,, $ where
\begin{equation}
\overline\rho_h[\,p^\mu]\,\equiv\,\left(\rho_h[\,p^\mu]\,\right)^\dagger
\Gamma^0\,,\quad \Gamma^0\,\equiv\,
\left[
\begin{array}{cc}
I & 0\\
0&-I
\end{array}
\right]\quad.
\end{equation}
The front-form $(1,0)\oplus(0,1)$ Majorana-like spinors, Table I,  should be
compared with the front-form $(1,0)\oplus(0,1)$ {\it Dirac-like} spinors
obtained in our recent work \cite{MS}. For instance,  in the massless limit for
the Dirac-like spinors,   the $h=\pm 1$ degrees of freedom are non-vanishing
and
the $h= 0$ degree of freedom identically vanishes. On the other hand, in the
massless limit, for the spin-$1$ Majorana-like spinors
$\rho^{S_\theta}[\,p^\mu]$, it is only the $h=+1$ degree of freedom that is
non-vanishing, while the $h=0$ and $h=-1$ degrees of freedom identically
vanish.

The origin of the above observation lies in the fact
\footnotemark[4]
\footnotetext[4]{Even though we make these observations for spin-$1$,
results similar to those
that follow are true for all spins (including spin-$1\over 2\,$).}
that the $(1,0)$ and $(0,1)$ boosts, $\Lambda_R(p^\mu)$ and $\Lambda_L(p^\mu)$,
 essentially become projectors of the $\phi^R_{+1}(p^\mu)$ and
$\phi^L_{-1}(p^\mu)$ as $m\to 0\,$.
To see this,  introduce
\begin{mathletters}
\begin{eqnarray}
{\cal Q}_R(m)&\,\equiv\,&
\left({m\over{p^+}}\right)\,\Lambda_R(p^\mu)\,=\
\left[
\begin{array}{ccc}
1 & 0 & 0\\
{\sqrt{2}}\,p^r/{p^+} & m/{p^+} & 0 \\
\left(p^r/p^+\right)^2 & \sqrt{2}\, m\, p^r /\left(p^+\right)^2 &
m^2/\left(p^+\right)^2
\end{array}
\right]\quad,\\
{\cal Q}_L(m)&\,\equiv\,&
\left({m\over{p^+}}\right)\,\Lambda_L(p^\mu)\,=\
\left[
\begin{array}{ccc}
m^2/\left(p^+\right)^2 & -\,\sqrt{2} \,m\, p_\ell /\left(p^+\right)^2
&\left(p^\ell/p^+\right)^2 \\
0 & m/p^+ & -\,\sqrt{2} \,p^\ell/p^+ \\
0&0&1
\end{array}
\right]\quad.
\end{eqnarray}
\end{mathletters}
The quasi-projector nature of ${\cal Q}_R(m\to 0)$ and
${\cal Q}_L(m\to 0)$ is immediately observed by verifying that:
${\cal Q}^2_R(m\to 0)={\cal Q}_R(m\to 0)$ and
${\cal Q}^2_L(m\to 0)={\cal Q}_L(m\to 0)\;$; but in general
${\cal Q}_R(m\to 0)+{\cal Q}_L(m\to 0) \not\to \openone\,$ and
${\cal Q}^\dagger_{R,L}(m\to 0)\not={\cal Q}_{R,L}(m\to 0)\,$.

To incorporate the $h=-1$ degree of freedom in the massless limit,
and to be able to treat the massive particles without introducing manifest
parity violation, we now repeat the above procedure for the
$\lambda^{S_\theta}[\,p^\mu]\,$
(and $\rho^{A_\theta}[\,p^\mu]$
and $\lambda^{A_\theta}[\,p^\mu]$  for the sake of completeness)
spinors. We find:
\begin{mathletters}
\begin{eqnarray}
&&\lambda_{-h}^{S_\theta}[\,p^\mu]\,=\,-(-1)^{|h|}\,
\rho_{h}^{S_\theta}[\,p^\mu]\quad,
\label{mmp}\\
&&\rho^{A_\theta}_h[\,p^\mu]\,=\,\Gamma^5\,\rho^{S_\theta}_h
[\,p^\mu]\,,\quad \Gamma^{5}\,=\,
\left[
\begin{array}{cc}
0 & \openone\\
\openone&0
\end{array}
\right]\quad,\label{extra}
\\
&&\lambda_{-h}^{A_\theta}[\,p^\mu]\,=\,(-1)^{|h|}\,
\Gamma^5\,\rho_{h}^{S_\theta}[\,p^\mu]\,=\,
(-1)^{|h|}\,\rho^{A_\theta}_h[\,p^\mu]
\quad;
\label{mmm}
\end{eqnarray}
\end{mathletters}
with $\openone=3\times 3$ identity matrix and
\begin{mathletters}
\begin{eqnarray}
&&\overline\rho^{\,S_\theta}_{h}[\,p^\mu]
\,\lambda^{S_\theta}_{h'}[\,p^\mu]\,=\, m^2\,\delta_{h h'}
\,=\,  \overline\rho^{\,A_\theta}_{h}[\,p^\mu]
\,\lambda^{A_\theta}_{h'}[\,p^\mu] \quad,\label{on}\\
&&\overline\rho^{\,A_\theta}_{h}[\,p^\mu]
\,\rho^{S_\theta}_{h'}[\,p^\mu]
\,=\, 0\,=\, \overline\rho^{\,S_\theta}_{h}[\,p^\mu]
\,\rho^{A_\theta}_{h'}[\,p^\mu] \quad.\label{onb}
\end{eqnarray} \end{mathletters}
As we will see in Sec. 6, the {\it bi-orthogonal} \cite{bio} nature of the
$\rho[\,p^\mu]$ and $\lambda[\,p^\mu]$ spinors results in a rather unusual
quantum field theoretic  structure for the $(1,0)\oplus(0,1)$ Majorana-like
field. Similar results hold true for other spins (including spin $1\over 2$).
The bi-orthogonal nature of the Majorana-like spinors is forced upon us by
self/antiself-$\theta$-conjugacy condition (\ref{sc})
 and cannot be changed as long as we require
that the basis-spinors correspond to  definite spin-projections (front-form
helicity-basis in our case).

It should now be
recalled that for the Dirac-like $(1,0)\oplus(0,1)$ spinors
$u_\sigma\{p^\mu\}$ and $v_\sigma\{p^\mu\}\,$, we know \cite{BWW} from the
associated wave equation that the $u_\sigma\{ p^\mu\}$ spinors are associated
with the forward-in-time propagating solutions (the ``positive energy
solutions'')  $u_\sigma\{ p^\mu\}\,\exp[-i(E\,t-{ \bf p\cdot x})]$, and
$v_\sigma\{ p^\mu\}$ spinors are associated with the backward-in-time
propagating
\footnotemark[5]
 \footnotetext[5] {Recall that the usual
interpretation of the ``negative energy'' states as  {\it anti}particles fails
(see p. 66 of Ref. \cite{BH}) for bosons. On the other hand, the
St\"uckelberg-Feynman framework \cite{SF} applies equally to fermions and
bosons.}  solutions (the ``negative energy solutions'')
$v_\sigma\{p^\mu\}\,\exp[+i(E\,t- {\bf p\cdot x})]$. Can one infer similar
results by studying the wave equation associated with the front-form
$(1,0)\oplus(0,1)$ Majorana-like spinors$\,$?

\noindent{\bf 5. Wave Equation for Majorana-Like $\protect\bbox{({\bf 1},0)
\oplus(0,{\bf 1})}$ Spinors}

Combining the Lorentz transformation properties for the $\phi_R(p^\mu)$ and
$\phi_L(p^\mu)$, given by Eqs. (\ref{r}) and (\ref{l}), with  the
definitions (\ref{fb}) of Majorana-like spinors,  we obtain the wave equations
satisfied by the $(1,0)\oplus(0,1)$ Majorana-like spinors. For the
$\rho\{p^\mu\}$ spinors, the wave equation we obtain reads (in chiral
representation, where it takes its simplest form):
\begin{equation}
\left[
\begin{array}{ccc}
-\zeta_\rho\,m^2\,\Theta_{[1]}&{\quad}&{\cal O}_1\\
{\cal O}_2&
{\quad}&-\zeta_\rho\,m^2\,\Theta_{[1]}
\end{array}
\right]
\rho\{p^\mu\}\,=\,0
\quad,
\label{majeq}
\end{equation}
where in the front form the operators   ${\cal O}_1$ and ${\cal O}_2$ are
defined as: $ {\cal O}_1\,=\, g_{\mu\nu}\,p^\mu\,p^\nu
\exp\left(-{\protect\bbox {\beta \cdot {\bf J}^\ast}}\right)\,\exp\left(
{\protect\bbox{ \beta^\ast\cdot  {\bf J}}}\right)\, $ and $ {\cal O}_2 =
g_{\mu\nu}\,p^\mu\,p^\nu \exp\left({\protect\bbox{ \beta^\ast\cdot {\bf
J}^\ast}}\right)\,\exp\left( - {\protect\bbox{\beta\cdot {\bf J}}}\right) \,.$
The non-zero elements of the front form (the
flat space time) metric $g_{\mu\nu}$
are: $g_{+-}={1\over 2}=g_{-+}$ and $ g_{11}=-1=g_{22}\,$. The wave equation
for the $\lambda\{p^\mu\}$ spinors is the same as Eq. (\ref{majeq})
 with $\zeta_\lambda$ being replaced by
$\zeta_\rho$.
The dispersion relations associated with the solutions of Eq.
(\ref{majeq}) are obtained by setting the determinant of the square bracket in
Eq. (\ref{majeq}) equal to zero. A simple, though somewhat lengthy, algebra
transforms the resulting equation into (true for all spin-$ 1$ Majorana-like
spinors, hence all reference to a specific spinor is dropped below): $
-\left(p^\ell\,p^r - p^+ p^- - \zeta\, m^2\right)^3\, \left(p^\ell\,p^r - p^+
p^- + \zeta\, m^2\right)^3 \,=\,0\,. $  As a result, the associated dispersion
relations read:
\begin{equation}
p^+\,=\,{ {p^\ell \,p^r\,+\zeta\, m^2}\over {p^-} }\,,\quad
p^+\,=\,{ {p^\ell \,p^r\,-\zeta \, m^2}\over {p^-} }\quad,
\end{equation}
each  with a multiplicity $3$ (for a {\it given} $\zeta$). Again, as seen in
Refs. \cite{DVAc,DVAf,SSB}, like  the case for the Dirac-like
$(1,0)\oplus(0,1)$ spinors, the wave equation for the Majorana-like
$(1,0)\oplus(0,1)$ spinors contains tachyonic degeneracy. For the Dirac-like
$(1,0)\oplus(0,1)$ spinors, we find that  the tachyonic solutions can be
reinterpreted as physical solutions within the context of a quartic self
interaction and spontaneous symmetry breaking \cite{SSB}.  Here, we concentrate
on the physically acceptable dispersion relations $p^+\,=\,(p^\ell \,p^r\,+
m^2)/ {p^-}$; or equivalently $E^2\,=\,{\bf p}^{\,2} \,+\, m^2\,$.

The wave equation satisfied by the plane  wave solutions $ \rho\{x\}
=\rho\{p^\mu\}\,\exp(-i\epsilon\,p^\mu x_\mu)\, $ and $ \lambda\{x\}
=\lambda\{p^\mu\}\,\exp(-i\epsilon\,p^\mu x_\mu)\, $ is obtained by first
expanding the exponentials in Eq. (\ref{majeq}), in accordance with the
identities given in Appendix A of Ref. \cite{MS}, and then letting
$p^\mu\rightarrow i\partial^\mu$. Next, to determine, $\epsilon$ we study the
resulting equation for the plane-wave solutions associated with  the rest
spinors. It is easily verified that for $\rho\{\,p^\mu\}$ as well as
$\lambda\{\,p^\mu\}$, it is not $\epsilon$ (directly)  but $\epsilon^2$ that is
constrained by the relation: $\epsilon^2= 1\,,$ giving $\epsilon=\pm 1\,$. This
is consistent with the intuitive understanding  in that we cannot distinguish
between the  forward-in-time propagating (``particles'')  and the
backward-in-time propagating (``antiparticles'') Majorana-like objects.
The above arguments are independent of which
representation we choose within the $(1,0)\oplus(0,1)$ representation space.

\noindent{\bf 6.
Majorana-Like $\protect\bbox{({\bf 1},0)
\oplus(0,{\bf 1})}$  Field Operator}

We now exploit the above considerations on the Majorana-like
spinors to construct the associated field operator. Generalizing the
spin-$1\over 2$ definition for a Majorana particle  of  Ref. \cite{BK},  we
define a general $(j,0)\oplus(0,j)$  Majorana-like field operator
$\Xi(x)$
\begin{equation}
U(C_\theta)\,\Xi(x)\,U^{-1}(C_\theta)
\,=\,\pm\,\Xi(x)
\quad. \label{u}
\end{equation}
In Eq. (\ref{u}), the ``$+$'' sign defines the self-$\theta$-conjugate and the
``$-$''  sign defines the antiself-$\theta$-conjugate field operator. The
explicit chiral-representation expression for $\theta$-conjugatiuon operator
$C_\theta$ as contained in Eq. (\ref{c}) is
\begin{equation}
C_\theta\,=\,{\bf C}_\theta\,K\,=\,\left[
\begin{array}{cc}
0&\xi\,\Theta_{[\,j]}\\
\left(\xi\,\Theta_{[\,j]}\right)^\ast&0
\end{array}
\right]\,K
\end{equation}
where $K$ complex conjugates (on the right) the objects in the Majorana-like
$(j,0)\oplus(0,j)$ representation space. For the example case of spin-$1$, when
Eqs. (\ref{u})  are coupled with the additional physical requirement that all
helicity degrees of freedom be treated symmetrically for manifest
$P$-covariance, the field operator $\Xi(x)$ is determined to be
\begin{mathletters}
\begin{eqnarray}
\Xi^{S_\theta}(x)  &=& \sum_{h=0,\pm1}\int {d^4 p}
{\Big[}{\cal
S}^{(\rho)}_{h}(p^\mu)\,\rho^{S_\theta}_{h}[\,p^\mu]\,\exp(-ip\cdot x)
\,+\,\eta_{_{GK}}\,
{\cal S}^{(\lambda)\,\dagger}
_h(p^\mu)\,\lambda^{S_\theta}_h[\,p^\mu]\,\exp(+ip\cdot x){\Big]}
\label{fos}\quad,\\
\Xi^{A_\theta}(x)  &=& \sum_{h=0,\pm1}\int {d^4 p}
{\Big[}{\cal A}^{(\rho)}
_{h}(p^\mu)\,\rho^{A_\theta}_{h}[\,p^\mu]\,\exp(-ip\cdot x)
\,+\,\eta_{_{GK}}\,
{\cal A}^{(\lambda)\,\dagger}
_h(p^\mu)\,\lambda^{A_\theta}_h[\,p^\mu]\,\exp(+ip\cdot
x){\Big]}\,,
\label{foa}
\end{eqnarray}
\end{mathletters}
where $\eta_{_{GK}}$ is the generalized
\footnotemark[6]
\footnotetext[6]{See footnote 19 of Ref. \cite{BK}.}
Goldhaber-Kayser phase factor;
and
\begin{mathletters}
\begin{eqnarray}
\left[{\cal S}^{(\rho)}_h(p^\mu)\,,\,\,{\cal S}^{(\rho)\,\dagger}
_{h'}(p^{\prime\,\mu})\right]&\,=\,&-\,(-1)^{|h|}\,
(2\pi)^3\,2 E(\vec p\,)\, \delta_{h,-h'}
\delta(\vec p- {\vec p}^{\,\prime})\quad,\label{coms}\\
\left[{\cal S}^{(\lambda)}_h(p^\mu)\,,\,\,{\cal S}^{(\lambda)\,\dagger}
_{h'}(p^{\prime\,\mu})\right]&\,=\,&-\,(-1)^{|h|}\,
(2\pi)^3\,2 E(\vec p\,)\, \delta_{h,-h'}
\delta(\vec p- {\vec p}^{\,\prime})\quad, \label{coma}
\end{eqnarray}
\end{mathletters}
with similar expression for the creation and annihilation operators of the
$\Xi^{A_\theta}(x)$ field. Several unusual features of expressions for the
field operators $\Xi(x)$, Eqs. (\ref{fos}) and (\ref{foa}), and commutators
(\ref{coms}) and  (\ref{coma}) should be explicitly noted:
\medskip
{ I.} The factor $-(-1)^{|h|} \,\delta_{h,-h'}$, rather than
the usual $\delta_{hh'}$, in the {\it r.h.s.}
 of Eqs. (\ref{coms}) and (\ref{coma})
arises from the bi-orthogonal \cite{bio} nature of $\rho[\,p^\mu]$ and
$\lambda[\,p^\mu]$ spinors.
{II.} The creation operator ${\cal S}^{(\lambda)\,\dagger}_h(p^\mu)$
for the plane wave $\lambda^{S_\theta}_h[p^\mu]\,\exp(ip\cdot x)$
is identical (within a
phase factor) to the
creation operator for the plane wave
 $\rho^{S_\theta}_{-h}[p^\mu]\,\exp(-ip\cdot x)$
\begin{equation}
{\cal S}^{(\lambda)\,\dagger}_h(p^\mu)\,=\,
-(-1)^{|h|}\,
{\cal S}^{(\rho)\,\dagger}_{-h}(p^\mu)\quad,
\end{equation}
with similar comments applicable to
${\cal A}^{(\lambda)\dagger}_h(p^\mu)$
and
${\cal A}^{(\rho)\,\dagger}_h(p^\mu)$.
\medskip

\noindent
In addition, in view of our results of Sec. 5,  the association of the
$\rho[\,p^\mu]$ spinors with the forward-in-time propagating solutions and
$\lambda[\,p^\mu]$ spinors with backward-in-time propagating solutions in the
explicit expressions of $\Xi(x)$ above is purely a convention.

Finally, we wish
to emphasize that the field operators we arrive at differ from similar
expressions \footnotemark[7] \footnotetext[7]{See, for example, Eq. (3.25) of
Ref. \cite{BK}; and  Eq. (2.5) of Gluza and  Zralek's paper in Ref.
\cite{Zralek}.} found in literature  for the $({1\over 2},0)\oplus(0,{1\over
2})$ Majorana field. Unlike the field operators $\Xi(x)$, these expressions
[even though they satisfy Eq. (\ref{u})] do not exploit the
Majorana-construction in the $(j,0)\oplus(0,j)$ representation space {\it and}
as a result cannot be expected to contain full physical content of a truly
neutral particle.

\noindent{\bf 7.
Concluding Remarks}

We have succeeded in  extending the
Majorana-construction for the $({1\over 2},0)\oplus(0,{1\over 2})$
representation space to all $(j,0) \oplus(0,j)$  representation spaces  despite
the general impression that Majorana's original construction was due to certain
``magic of Pauli matrices.'' We studied the $(1,0)\oplus(0,1)$ Majorana-like
representation space  in some detail and presented an associated wave equation.
Since nature has a host of neutral ``fundamental particles'' of spin-$ 1$ and
-$2$ and composite hadronic structures of even higher spins, the existence of
the Majorana-like $(j,0)\oplus(0,j)$ representation spaces introduced in this
work may have some physical relevance for the unification beyond the
electroweak theory and hadronic phenomenologies. In the massless limit,
$(j,0)\oplus(0,j)$ fields, independent of spin and independent of whether they
are Dirac-like or Majorana-like,  contain only two helicity degrees of freedom.
This observation allows the construction of higher-spin field theories without
introducing or imposing any auxiliary fields, negative-norm states, or
constraints. This fact may have some significance for theories involving
supersymmetric transformations, which transform between fermions and bosons,
and which are normally rife with non-physical, additional fields. It should be
explicitly noted that even though the construction of the $(j,0)\oplus(0,j)$
Majorana-like fields is {\it manifestly} covariant under parity, in general
massive Majorana-like particles carry {\it imaginary} intrinsic parity
\cite{Parity} and hence these particles in interactions with
Dirac-like/Dirac particles (like charged leptons and quarks) naturally lead to
non-conservation of parity.

\noindent{\bf
Acknowledgments}
One of the authors (D.V.A.) warmly thanks Boris Kayser for a conversation, and
Otto Nachtmann and Marek Zralek for correspondence, on the  subject of Majorana
fields.

\narrowtext
\begin{table}
\caption{Spin-$ 1$ self-$\theta$-conjugate Majorana-like
$\rho^{S_\theta}[\,p^\mu]$ spinors. Here $p^\pm=E\,\pm\,p_z$,
$p^r=p_x\,+\,i\,p_y$ and $p^\ell=p_x\,-\,i\,p_y$. The subscript $h=0,\,\pm 1$
on $\rho^{S_\theta}_h[\,p^\mu]$ refers to the {\it front-form helicity}
\protect\cite{MS} degree of freedom. The remaining spinors
$\lambda^{S_\theta}[\,p^\mu]$, $\rho^{A_\theta}[\,p^\mu]$ and
$\lambda^{A_\theta}[\,p^\mu]$ are related to  $\rho^{S_\theta}[\,p^\mu]$ via
Eqs. (\protect\ref{mmp}) to (\protect\ref{mmm}).}
\begin{tabular}{ccc}
$\rho^{S_\theta}_{+1}[\,p^\mu]$
&$\rho^{S_\theta}_{0}[\,p^\mu]$
&$\rho^{S_\theta}_{-1}[\,p^\mu]$\\
\tableline
${1\over 2}
\left[
\begin{array}{c}
p^+\,+\,({p^\ell}^2/p^+)\\
\sqrt{2}\,(p^r\,-\,p^\ell) \\
p^+\,+\,({p^r}^2/p^+)\\
p^+\,-\,({p^\ell}^2/p^+)\\
\sqrt{2}\,(p^r\,+\,p^\ell) \\
-\,p^+\,+\,({p^r}^2/p^+)
\end{array}
\right]$ &
${m \over\sqrt{ 2}}
\left[
\begin{array}{c}
p^\ell/p^+\\
0\\
p^r/p^+\\
-\,p^\ell/p^+ \\
\sqrt{2} \\
p^r/p^+
\end{array}
\right]$&
$\,{m^2 \over 2}
\left[
\begin{array}{c}
1/p^+\\
0\\
1/p^+\\
-\,1/p^+ \\
0\\
1/p^+
\end{array}
\right]$\\
\end{tabular}
\end{table}
\widetext

 \end{document}